\documentclass[11pt]{article}
\usepackage{epsfig}

\newcommand{\be}{\begin{eqnarray}}
\newcommand{\ee}{\end{eqnarray}}

\def\ll#1{\left#1}
\def\r#1{\right#1}
\def\fr{\frac{1}{2}}

\def\mref#1{(\ref{#1})}

\def\p{\partial}

\def\bd{\begin{displaymath}}
\def\ed{\end{displaymath}}

\def\ba#1{\begin{array}{#1}}
\def\ea{\end{array}}
\def\nn{\nonumber}

\begin{document}

\pagestyle{empty}

\begin{center}

{\LARGE\bf  Space -- time symmetry \\of noncommutative field theory}
\vskip 18pt

{\large {\bf Cezary Gonera{$\dag$}, Piotr Kosi\'nski{$\dag$},  Pawe{\l}  Ma\'slanka{$\dag$} and
Stefan Giller{$\ddag$}}}

\vskip 3pt

{$\dag$}Department of Theoretical Physics II, University of {\L}\'od\'z,\\
ul. Pomorska 149/153, 90-236 {\L}\'od\'z, Poland\\
e-mail: cgonera@uni.lodz.pl, pkosinsk@uni.lodz.pl, pmaslan@uni.lodz.pl\\
$\ddag$ Institute of Physics, Jan Dlugosz Academy in Czestochowa\\
ul. Armii Krajowej 13/15, 42-200 Czestochowa\\
e-mail: sgiller@uni.lodz.pl
\end{center}
\vspace{6pt}
\begin{abstract}
We consider the deformed Poincare group describing the space-time symmetry of noncommutative
field theory. It is shown how the deformed symmetry is related to the explicit symmetry breaking.
\end{abstract}
\vskip 9pt

{\small PACS number(s): 11.10-z, 11.30Cp}


\newpage

\pagestyle{plain}

\setcounter{page}{1}

\section*{1. Introduction}

\hspace{10pt} In the two recent interesting papers \cite{1,2} (cf. also \cite{3})
Chaichian et al.
proposed a new interpretation of the  symmetry of noncommutative space-time defined by the commutation
relations:
\be
[{\hat x}^\mu,{\hat x}^\nu]=i\Theta ^{\mu\nu}
\label{1}
\ee
where $\Theta ^{\mu\nu}$ is a constant antisymmetric matrix.

According to the standard wisdom the relation \mref1 break Lorentz symmetry down to
the stability subgroup of $\Theta ^{\mu\nu}$. Therefore, the exact symmetry of noncommutative
space-time is the semidirect product of the latter and translation group. In spite of
 that all fundamental issues of the noncommutative quantum field theory (NCQFT)
 \cite{4,5} are discussed in fully covariant approach using the representations of
the Poincare group. Moreover, when trying to base the theory on the stability subgroup of
$\Theta ^{\mu\nu}$ one is at once faced with the question why the multiplets of stability
subgroup are organised in such a way as to form the complete multiplets of full
Lorentz/Poincare group.

The way out of this dilemma proposed by Chaichian et al. consists in the following.
One poses the question whether the noncommutative space-time admits as large symmetry
as its commutative counterpart provided the symmetry is understood in the wider sense
of quantum groups theory. If this is the case one can try to use this generalised symmetry
as the substitute of Poincare group.

Chaichian et al. have shown that this scenario works quite well. Actually,
they constructed the infinitesimal version of deformed Poincare symmetry which
appeared to be the standard Poincare algebra equipped, however, with the modified
(twisted) coproduct. They gave also a number of convincing  arguments that the
generalised symmetry is as efficient as the Poincare symmetry in the commutative case.

On the other hand one can consider NCQFT from different point of view. It emerges,
in particular circumstances, as a specific limit of fully symmetric theory. Therefore,
it is not simply an example of the theory with some prescribed symmetry group which,
accidentally, appears to be a subgroup of the Poincare group. On the contrary, it is
an example of what is called explicit symmetry breaking.

Explicit symmetry breaking is a well-known notion. Generically such a symmetry
pattern can be described as follows. All dynamical variables as well as all
parameters entering have well-defined transformation properties under some group $G$.
However, $G$ itself is not good candidate for the symmetry group except that all
{\it parameters} are invariant under $G$. If this is not the case only the
stability subgroup $H\subset G$ of all parameters can serve as a symmetry group.

On the lagrangian level all that means that the Lagrangian is invariant under
$G$-trans\-for\-ma\-tion of all dynamical variables as well as all parameters. However,
symmetry demands more, namely the invariance under the transformation of dynamical
variables provided the values of parameters are kept fixed. Consequently symmetry
transformations must belong to $H$.

When the symmetry group $H$ emerges from explicit breaking of the larger group $G$
one can say more about the structure of the system than it follows by merely viewing
$H$ as a symmetry group. Some aspects can be explained in terms of the exact symmetry
$H$ while in order to understand other properties one has to appeal to the initial
group $G$.  On the Lagrangian level one
expects, for example, the currents related to the generators of $H$ to be conserved while
those related to generators of $G/H$ not; however,the invariance of the lagrangian under simultaneous
 $G$-transformations of dynamical variables and parameters restricts the admissible form of current divergencies.
 In fact, the (nonvanishing) divergencies can be
calculated from the variation of the Lagrangian when the parameters vary under the
action of $G$.

The point we want to make in the present note is that in some cases the explicit
symmetry breaking $G\downarrow H$ can be described in terms of deformed $G$-symmetry,
the deformation parameters being related to the parameters appearing in those terms of the lagrangian
which are responsible for symmetry breaking..
The quantum symmetry, being a more general notion, should impose less severe
restrictions on basic characteric of the system. On the other hand, deformed $G$
(containing undeformed $H$ as a substructure) is more than $H$ itself. Therefore we
cannot expect quantum $G$ to imply conservation of all currents but we can expect it to
impose some restrictions on current divergencies; in fact, they should be the same as those
implied by explicit $G\downarrow H$ breaking. We show that this scheme really works
in the case of field theory on noncommutative space-time. To this end we consider the
global counterpart of infinitesimal group of Chaichian et al. and discuss
the Noether identities following from generalised symmetry. They are the same as the
ones implied by explicit symmetry breaking and the divergencies of the
corresponding currents can be related equivalently either to the variation of the
Lagrangian under the change of the symmetry breaking parameters or deformation of the
symmetry group.

\section*{2. Global $\Theta$-Poincare symmetry}

\hskip+1.5em The global counterpart of Chaichian et al. algebra has been found few years ago \cite{6,7}
and rediscovered recently \cite{8}.

The corresponding *-Hopf algebra is generated by Hermitean elements $\hat\Lambda^\mu\;_\nu,\;\hat a^\mu$ obeying
\be
\ba l
[{\hat a}^\mu,{\hat a}^\nu]=-i\Theta ^{\rho\sigma}(\hat\Lambda^\mu\;_\rho\hat\Lambda^\nu\;_\sigma-\delta^\mu\;_\rho
\delta^\nu\;_\sigma)\\
\ll[\hat\Lambda^\mu\;_\nu,\bullet\r] =0   \\
\Delta(\hat a^\mu)=\hat\Lambda^\mu\;_\nu\otimes\hat a^\nu+\hat a^\mu\otimes{\bf 1}\\
\Delta(\hat\Lambda^\mu\;_\nu)=\hat\Lambda^\mu\;_\alpha\otimes \hat\Lambda^\alpha\;_\nu      \\
S(\hat a^\mu)=-\hat\Lambda_\nu\;^\mu\hat a^\nu \\
S(\hat\Lambda^\mu\;_\nu)=\hat\Lambda_\nu\;^\mu\\
\epsilon(\hat a^\mu)=0\\
\epsilon(\hat\Lambda^\mu\;_\nu)=\delta^\mu\;_\nu
\ea
\label{2}
\ee

The relations \mref2 define a deformation $P_{\Theta}$ of Poincare group ($\Theta$ -Poincare group). $P_{\Theta}$ acts on
noncommutative space-time in the standard way:
\be
\hat x^\mu \rightarrow  \hat\Lambda^\mu\;_\nu\otimes\hat x^\nu+\hat a^\mu\otimes{\bf 1}
\label{3}
\ee

Let us note that the stability subgroup of $\Theta^{\mu\nu}$ is a classical group.

The noncommutative algebraic structure can be encoded in commutative framework via Weyl-Moyal correspondence. Let $M_\Theta$ be
the algebra spanned by the generators $\hat x^\mu$ ($\Theta$-Minkowski space-time). Then $P_{\Theta}\otimes M_\Theta$
has a very simple structure because Lorentz elements $\hat\Lambda^\mu\;_\nu$ belong to its center and can be viewed as
commutative parameters. Writing eq.\mref2 in the form $[{\hat a}^\mu,{\hat a}^\nu]=i\tilde\Theta^{\mu\nu}$ with
$\tilde\Theta^{\mu\nu}=-\Theta ^{\rho\sigma}(\hat\Lambda^\mu\;_\rho\hat\Lambda^\nu\;_\sigma-\delta^\mu\;_\rho
\delta^\nu\;_\sigma)$ one can use Weyl-Moyal correspondence for the algebra spanned by $\hat x^\mu$ and $\hat a^\mu$.
As a result we obtain the following star product (see \cite{9} for the idea to use the star product for the algebra
of functions depending on group elements):
\be
&&(f*g)(\Lambda,a,x)\equiv\nn\\
&&f(\Lambda,a,x)\exp\ll[\frac{i}{2}\ll(\Theta^{\mu\nu}\frac{\stackrel{\leftarrow}{\p}}{\p x^\mu}
\frac{\stackrel{\rightarrow}{\p}}{\p x^\nu}-\Theta ^{\rho\sigma}(\Lambda^\mu\;_\rho\Lambda^\nu\;_\sigma-\delta^\mu_\rho
\delta^\nu_\sigma)\frac{\stackrel{\leftarrow}{\p}}{\p a^\mu}\frac{\stackrel{\rightarrow}{\p}}{\p a^\nu}\r)\r]g(\Lambda,a,x)
\label{4}
\ee
for any two functions $f,\;g$ of {\it classical} variables $\Lambda^\mu\;_\nu,\;x^\mu,\;a^\mu$.

The above star product is by construction associative. For $f,\;g$ depending on $x^\mu$ only
one obtains standard Moyal product while for the functions on classical Poincare group manifold our product encodes the
algebraic structure defined by first two eqs.\mref2.

The crucial property of our star product is that it commutes with the action of the Poincare group:
\be
\ll.f(x)*g(x)\r|_{x\to\Lambda x+a}=f(\Lambda x+a)*g(\Lambda x+a)
\label{5}
\ee

On the LHS one takes first the star product ( the standard Moyal one ) and then applies Poincare transformation while
on the RHS Poincare transformation is taken before applying the star product \mref4.

Eq.\mref5  can be easily proven by noting that:
\be
\frac{\p f(\Lambda x+a)}{\p x^\mu}=\ll.\Lambda^\nu\;_\mu\frac{\p f(x)}{\p x^\nu}\r|_{x\to\Lambda x+a}\nn\\
\frac{\p f(\Lambda x+a)}{\p a^\mu}=\ll.\frac{\p f(x)}{\p x^\mu}\r|_{x\to\Lambda x+a}
\label{6}
\ee

It is the global counterpart of the infinitesimal form obtained by Chaichian et al. (see \cite{2} esp.
Appendix). This local form can be derived as follows. One defines the Lie algebra generators by standard duality
rules, which in the case of Poincare group read
 ($\Delta^\mu\;_\nu\equiv \Lambda^\mu\;_\nu-\delta^\mu\;_\nu$), (cf.\cite{10}):
\be
<P_\mu\;,\;a^{\nu_1}...a^{\nu_m}\Delta^{\rho_1}\;_{\sigma_1}...\Delta^{\rho_n}\;_{\sigma_n}>=
i\delta_{m1}\delta_{n0}\delta_\mu\;^{\nu_1}\nn\\
<M_{\alpha\beta}\;,\;a^{\nu_1}...a^{\nu_m}\Delta^{\rho_1}\;_{\sigma_1}...\Delta^{\rho_n}\;_{\sigma_n}>=
i\delta_{m0}\delta_{n1}(\delta_\alpha^{\rho_1}g_{\beta\sigma_1}-\delta_\beta^{\rho_1}g_{\alpha\sigma_1})
\label{7}
\ee

The action of the generators is defined in the standard way \cite{11}: let $g$ be an element of Hopf algebra,
$\chi$ - an element of dual Hopf algebra and $gx=\sum_ig_i\otimes x_i$ - the group action; then the action of
dual algebra is defined by $\tilde\chi x=\sum_i<\chi,g_i>x_i$. Using this together with eqs. \mref4, \mref5 and \mref7
one easily finds
\be
\begin{array}{rcl}
{\tilde P}_\mu f&=&i\p_\mu f\\
{\tilde M}_{\alpha\beta}f&=&-i(x_\alpha\p_\beta-x_\beta\p_\alpha)f\\
{\tilde P}_\mu(f*g)&=&({\tilde P}_\mu f)*g+f*({\tilde P}_\mu g)\\
{\tilde M}_{\alpha\beta}(f*g)&=&({\tilde M}_{\alpha\beta}f)*g+f*({\tilde M}_{\alpha\beta}g)+
\fr\Theta^{\mu\nu}\ll(\eta_{\mu\alpha}({\tilde P}_\beta f)*({\tilde P}_\nu g)-\r.\\
               &&\ll.\eta_{\mu\beta}({\tilde P}_\alpha f)*({\tilde P}_\nu g)+
                     \eta_{\nu\alpha}({\tilde P}_\mu  f)*({\tilde P}_\beta  g)-
                     \eta_{\nu\beta}({\tilde P}_\mu  f)*({\tilde P}_\alpha g)\r)
\ea
\label{8}
\ee

The last of eq.\mref8 coincides with eq.(2.8) of Ref.\cite{1} (see also \cite{12}). This concludes our discussion
of the global $\Theta$-Poincare symmetry. Other details concerning the geometry of $P_\Theta$ will be given elsewhere.

\section*{3. $\Theta$-Poincare symmetry vs. explicit symmetry}

\hspace{20pt}Let us now consider the noncommutative field theory in the Lagrangian formalism. Assume for simplicity
that we are dealing with scalar fields only; the action of Poincare group is given by:
\be
\Phi(x)\to\Phi(\Lambda x+a)
\label{9}
\ee
which differs from the standard action ($\Phi(x)\to\Phi(\Lambda^{-1}(x-a))$) but this is irrelevant for our purposes.

The general form of the Lagrangian reads:
\be
L(x)=L(\Phi(x),\p_\mu\Phi(x))_*
\label{10}
\ee
where the star on the right-hand side means that all products are the star ones (although some of them can be replaced
by normal products when $L$ is integrated to yield the action).

$L(x)$ can be viewed as a standard Lagrangian although depending on derivatives of arbitrary orders:
\be
L(x)=\bar L\ll(\Phi(x),\p_\mu\Phi(x),\p_\mu\p_\nu\Phi(x),...;\Theta^{\alpha\beta}\r)
\label{11}
\ee
Here we have written out explicitly its dependence on the parameters $\Theta^{\alpha\beta}$.

Now, using \mref{11} as a starting point one can derive Noether's identities. It is slightly complicated due to the
appearance of higher-order derivatives but proceeds along the standard way. One writes out the condition of invariance
of the Lagrangian under a given transformation (Lorentz one in the case under consideration) and attempts to rearrange it
as to obtain the Euler-Lagrange expressions plus total divergence of something (the corresponding current). In our case
the initial identity contains an additional term $\delta\Theta^{\mu\nu}\frac{\p\bar L}{\p\Theta^{\mu\nu}}$ because
$\bar L$ is Lorentz invariant only provided $\Theta^{\mu\nu}$ transforms as an antisymmetric tensor of second rank.
This term gives rise to the nonconservation of the four dimensional angular momentum. Obviously the divergence of the
relevant current is defined only up to the terms proportional to Euler-Lagrange expressions and the terms of the form
$\p_\mu X^\mu$, $X^\mu$ being an arbitrary four-vector constructed out of the fields and their derivatives. In
particular if it can be written as a sum of such terms the corresponding current can be redefined to yield the
conserved one.

The term $\delta\Theta^{\mu\nu}\frac{\p\bar L}{\p\Theta^{\mu\nu}}$ can be easily calculated by noting that
$\Theta^{\mu\nu}$ enters $L(x)$ only through the star product. Under the variation $\Theta^{\mu\nu}\to\Theta^{\mu\nu}+
\delta\Theta^{\mu\nu}$ the star product transforms as follows:
\be
f*g\to f*g+\frac{i}{2}\delta\Theta^{\mu\nu}\p_\mu f*\p_\nu g
\label{12}
\ee
Let us consider the Lorentz transformation in the $(\alpha \beta )$-plane, $\Lambda ^\mu _{\;\;\nu }= (exp(\frac{\omega }
{2}M_{\alpha \beta }))^\mu _{\;\;\nu }$\ with $(M_{\alpha \beta })^\mu _{\;\;\nu }\equiv \delta _{\;\;\alpha} ^{\mu }
g_{\beta \nu }
-\delta _{\;\;\beta }^{\mu }g_{\alpha \nu }$\ being the corresponding generator in the defining representation; infinitesimally,
$\Lambda ^\mu _{\;\;\nu }\simeq \delta ^{\mu} _{\;\;\nu }+\frac{\omega }{2}(\delta ^{\mu }_{\;\;\alpha }g_{\beta \nu }-
\delta ^{\mu }_{\;\;\beta }g_{\alpha \nu })$.
Inserting into eq.(12) the form of $\delta\Theta^{\mu\nu}$ induced by this transformation:
\be
\delta\Theta^{\mu\nu}_{(\alpha\beta)}=\frac{\omega}{2}
(\delta_\alpha^{\mu}g_{\beta\rho}-\delta_\beta^{\mu}g_{\alpha\rho})\Theta^{\rho\nu}+\nn\\
+\frac{\omega}{2}(\delta_\alpha^{\nu}g_{\beta\rho}-\delta_\beta^{\nu}g_{\alpha\rho})\Theta^{\mu \rho }
\label{13}
\ee
one gets ($\Delta_{\alpha\beta}^{\;\;\;\;\;\mu}\equiv\Theta_\alpha\;^\mu\p_\beta-\Theta_\beta\;^\mu\p_\alpha$):
\be
f*g\to f*g+\frac{i\omega}{4}\Delta_{\beta\alpha}^{\;\;\;\;\;\nu}f*\p_\nu g -
\frac{i\omega}{4}\p_\mu f*\Delta_{\beta\alpha}^{\;\;\;\;\;\mu} g
\label{14}
\ee

Note the similarity between eq.\mref{14} and the last eq.\mref8.

If $f$ and/or $g$ are themselves the star products of other factors one has to apply \mref{14} successively
keeping only terms of first order in $\omega$. As a result one obtains for the variation of the
product of arbitrary number of factors the sum of terms corresponding to all pairs of factors:
\be
&&\delta(f_1*f_2*...*f_n)=\nn\\
&&\frac{i\omega}{4}(\Delta_{\beta\alpha}^{\nu\mu}-\Delta_{\beta\alpha}^{\mu\nu})
(\sum_{k<l}f_1*...*\p_\mu f_k*...*\p_\nu  f_l*...*f_n )
\label{15}
\ee
where $\Delta_{\alpha\beta}^{\mu\nu}\equiv\Theta_\alpha\;^\mu\delta_\beta^\nu-\Theta_\beta\;^\nu\delta_\alpha^\mu$.

Eq.\mref{15} allows us to calculate the divergence of Lorentz current for any theory. Of course, as we have mentioned
above, the resulting expression can be further modified by adding/subtracting admissible terms.

Let us consider now the Noether identities within $\Theta$-Poincare symmetry formalism. In the nondeformed case the
equivalent way of deriving the Noether identities is to compute the action of Lorentz generator on $L$. Adopting this
method in $\Theta$-Poincare case we apply $i\tilde M_{\alpha\beta}$ to $L$ and use eq.\mref8. Now, $L$ is a sum (finite or
infinite) of star-product monomials in fields and their derivatives so one has only to compute the action of
$i\tilde M_{\alpha\beta}$ on such monomials. According to standard arguments \cite{11} this amounts to find
$...({id}\otimes{id}\otimes\Delta)\circ({id}\otimes\Delta)\circ\Delta\tilde M_{\alpha\beta}$. Using eq.\mref8 we finally
arrive at the expression of the form:
\be
i\tilde M_{\alpha\beta}(f_1*f_2*...*f_n)=\sum_{k=1}^nf_1*...*i\tilde M_{\alpha\beta}f_k*...*f_n+\nn\\
\frac{i}{2}\sum_{k<l}\ll(f_1*...*(\Theta_\alpha\;^\nu\tilde P_\beta-\Theta_\beta\;^\nu\tilde P_\alpha)
f_k*...*P_\nu f_l*...*f_n+\r.\nn\\
\ll.-f_1*...*P_\nu f_k*...*(\Theta_\alpha\;^\nu\tilde P_\beta-\Theta_\beta\;^\nu\tilde P_\alpha) f_l*...*f_n\r)
\label{16}
\ee
which is equivalent to eq.\mref{15}.

When calculating $i\tilde M_{\alpha\beta}L$ we get two types of terms generated by two sums on the right-hand side of
eq.\mref{16}. First there will be a sum of terms corresponding to various monomials entering the Lagrangian with the operator
$i\tilde M_{\alpha\beta}$ inserted in all possible ways. This is one-to-one correspondence with the commutative case
except that the proper ordering must be observed due to noncommutativity of the star product. In order to use field
equations one has to rearrange the ordering as to put the terms with $i\tilde M_{\alpha\beta}$ left- or rightmost.
This yield some star commutators; however, it is known that such commutators can be written as total divergence and
included in the divergence of the angular momentum current. We conclude that the angular momentum tensor can be
obtained by properly ordering its commutative counterpart, replacing ordinary products by the star ones and adding a number
of well-defined expressions following from star-commutators which appear due to reordering.

The second sum on the right-hand side of eq.\mref{16} gives rise to the divergence of the angular-momentum current. It
is exactly the same term as the one obtained from explicit symmetry breaking.

Let us conclude with a simple example $\Phi^3$-theory (see Refs. \cite{13}$\div$ \cite{16} for similar examples):
\be
L=\fr\ll(\p_\mu\Phi*\p^\mu\Phi-m^2\Phi*\Phi\r)-\frac{\lambda}{3!}\Phi^3_*
\label{17}
\ee

The action $S=\int d^4xL$\ is invariant under translations; applying standard Noether algorithm ( which is equivalent
to computing $\partial _\mu L$) 
 one arrives at the following identity:
\be
\p_\mu\ll(\fr(\p_\alpha\Phi*\p^\mu\Phi+\p^\mu\Phi*\p_\alpha\Phi)-\delta_\alpha^\mu L\r)-
\fr\ll((\p^2+m^2)\Phi+\frac{\lambda}{2!}\Phi^2_*\r)*\p_\alpha\Phi+\nn\\
-\fr\p_\alpha\Phi*\ll((\p^2+m^2)\Phi+\frac{\lambda}{2!}\Phi^2_*\r)-\frac{\lambda}{2\cdot 3!}
\ll[\Phi,[\p_\alpha\Phi,\Phi]_*\r]_*=0
\label{18}
\ee

Now the $\star$-commutator is a total divergence:
\be
-\frac{\lambda}{2\cdot 3!}\ll[\Phi,[\p_\alpha\Phi,\Phi]_*\r]_*\equiv \p_\mu Y_\alpha\;^\mu
\label{19}
\ee

Therefore, the "corrected" energy-momentum tensor:
\be
T_{*\alpha}\;^\mu\equiv \fr(\p_\alpha\Phi*\p^\mu\Phi+\p^\mu\Phi*\p_\alpha\Phi)-\delta_\alpha^\mu L+ Y_\alpha\;^\mu
\label{20}
\ee
is conserved, as it should be since the theory is translationally invariant.

On the other hand, proceeding as described above and making some rearrangement of terms we find the following identity
for Lorentz transormations:
\be
&&\p_\mu\ll(x_\alpha\ll(\fr(\p_\beta\Phi*\p^\mu\Phi+\p^\mu\Phi*\p_\beta\Phi-\delta_\beta^\mu L\r)\r.+\nn\\
&&\ll.-x_\beta\ll(\fr(\p_\alpha\Phi*\p^\mu\Phi+\p^\mu\Phi*\p_\alpha\Phi-\delta_\alpha^\mu L\r)-
\frac{\lambda}{2\cdot 3!}\p^{-1}_\mu\ll[[\Phi,i\tilde M_{\alpha\beta}\Phi]_*,\Phi\r]_*\r)+\nn\\
&&-\fr\ll((\p^2+m^2)\Phi+\frac{\lambda}{2!}\Phi^2_*\r)*i\tilde M_{\alpha\beta}\Phi
-\fr i\tilde M_{\alpha\beta}\Phi*\ll((\p^2+m^2)\Phi+\frac{\lambda}{2!}\Phi^2_*\r)+\nn\\
&&+\frac{i}{4}\Delta_{\alpha\beta}^\nu\Phi*\p_\nu\ll((\p^2+m^2)\Phi+\frac{\lambda}{2!}\Phi^2_*\r)+
\frac{i}{4}\p_\nu\ll((\p^2+m^2)\Phi+\frac{\lambda}{2!}\Phi^2_*\r)*\Delta_{\alpha\beta}^\nu\Phi +\nn\\
&&-\frac{\lambda}{3!}\ll(\frac{i}{2}(\p_\nu\Phi*\Delta_{\alpha\beta}^\nu\Phi*\Phi-
\Phi*\Delta_{\alpha\beta}^\nu\Phi*\p_\nu\Phi)+
\frac{i}{4}\ll[\Delta_{\alpha\beta}^\nu\Phi,\lbrace\Phi,\p_\nu\Phi\rbrace_*\r]_*\r)=0
\label{21}
\ee

One obtains the same result  by computing the divergence of $x_\alpha T_{*\beta}\;^\mu-x_\beta T_{*\alpha}\;^\mu$ 
using eq.\mref{20}.

\section*{4. Concluding remarks}

\hspace{20pt}We have shown that $\Theta$-Poincare group defined in Refs. \cite{7}, \cite{8} provides 
the space-time symmetry
group for noncommutative field theory. It gives rise to proper Noether identities modified as compared with the
commutative case by terms following from the modification of algebraic (in global formulation) or coalgebraic (in
infinitesimal version) structure. In the case under consideration we arrive at the clear understanding of the meaning
of deformed symmetry. It emerges due to the fact that we are considering the theory with explicit symmetry breaking. 
This means that, although the dynamics (lagrangian) is not invariant under the full Poincare group, all dynamical 
variables as well as parameters entering have well-defined transformation properties. The symmetry breaking
 manifests itself in the form of current nonconservation which is computed using the invariance of the lagrangian
 under simultaneous transformations of dynamical variables and parameters. These properties can be summarized as 
resulting from deformed Poincare symmetry. In particular, the deformation is responsible for the nonconservation 
of some currents in spite of the fact that the deformed symmetry can be viewed as the exact one.

We have considered very special example of such an equivalence scheme. The symmetry breaking emerges here through the
modification of the product of basic dynamical variables. One can ask whether there are other examples of the
complementarity between explicit symmetry breaking picture and the one based on
deformed exact symmetry and whether such a deformation is always given
by a twist.

Finally, we note that the problem of symmetries of noncommutative theories with Galilean invariance has been considered in
\cite{17}, \cite{18}. 

\section*{Acknowledgements}

\hspace{20pt}This paper has been supported by the grant 1P03B02128 of Polish Ministry of Science

\end{document}